\documentclass{article}
\usepackage{geometry}
 \usepackage[T1]{fontenc}
 \usepackage{float}

\usepackage{comment}

\usepackage{authblk}
\usepackage{graphicx} 
\usepackage{subfigure}
\usepackage{caption}
\usepackage{subcaption}
\usepackage{mwe}
\usepackage[T1]{fontenc}

\usepackage[linktoc=all,colorlinks=true,citecolor=blue,linkcolor=blue]{hyperref}

\newcommand{\subf}[2]{%
  {\small\begin{tabular}[t]{@{}c@{}}
  #1\\#2
  \end{tabular}}%
}

\title{Indicator Functions Detect Tangentially Transient Behaviour on
Decaying Normally Hyperbolic Invariant Manifolds}

\author[1,2,3]{Francisco Gonzalez Montoya \thanks{ f.gonzalezmontoya@leeds.ac.uk, \\f.gonzalez.montoya@ciencias.unam.mx}}

\author[2]{ Christof Jung }

\small{

\affil[1]{  Faculty of  Engineering and Physical Sciences, University of Leeds, Leeds, LS2 9JT, 
 United Kingdom}

\affil[2]{Instituto de Ciencias Físicas, Universidad Nacional Aut\'onoma de M\'exico, Av. Universidad s/n, Col.
Chamilpa, Cuernavaca, Morelos, CP 62210, México }

\affil[3]{Facultad de Ciencias,  Universidad Nacional Aut\'onoma de M\'exico, Av. Universidad 3000,
Circuito Exterior s/n, Coyoacán, CP 04510, Ciudad Universitaria, Ciudad de M\'exico, México}

}

\date{\today}

\begin{document}

\maketitle

\begin{abstract}
We study the decay scenario of a codimension-2 NHIM in a three-degrees-of-freedom Hamiltonian system under increasing perturbation when the NHIM loses
its normal hyperbolicity. On one hand, we follow this decay in the Poincar\'e
map for the internal dynamics of the NHIM. On the other hand, we also
follow the decay in a time delay function calculated on a 2-dimensional plane in the phase space of the
system. In addition, we observe the role of tangential transient effects on the decaying NHIM
and their manifestation in the delay time indicator function. We obtain ideas on how the decay of
NHIMs and the tangential transient effects are encoded in indicator functions. As an example
of demonstration, we use the motion of an electron in a perturbed magnetic dipole field.
\end{abstract}

\section{Introduction}

The global dynamics of a chaotic dynamical system is governed to a large extent by
its unstable invariant subsets in the phase space, in particular by the ones of codimension 2.
These subsets have stable and unstable manifolds of codimension 1 which can form walls
and tubes in the phase space that direct the general dynamics. These considerations
of codimensions hold equally well for the constant energy manifold of Hamiltonian flows as well as for the
corresponding Poincar\'e maps. A remarkable kind of invariant sets in the phase space are
normally unstable invariant surfaces known under the name ``Normally Hyperbolic Invariant
Manifolds'', standard abbreviation NHIMs. For general properties of these manifolds see
\cite{wiggins}.

For any important subsets in a phase space, the question of their stability always arises under general perturbations of the system. For NHIMs, the following persistence theorem holds:
NHIMs survive perturbations and keep their global topology as long as their normal instability
remains larger than their tangential instability. This property is called ``normal hyperbolicity''.  Several methods to prove the persistence theorem of
NHIMs are developed in the references \cite{wiggins,fenichel,berger,eldering}.
Examples indicate that NHIMs start to decay locally as soon as in some
point on the NHIM the tangential instability becomes larger than the normal instability.
For three instructive examples see Figs. 4 and 5 and their explanation
in \cite{fenichel}, example 1.1 with Fig.1.3 in subsection 1.2.1 of \cite{eldering}, and
Fig.9 and its explanation in section 6 of \cite{gonzalez2}.

The persistence theorem does not imply the structural stability of the internal dynamics of the NHIMs.
Under general perturbations, we find qualitative changes in the internal dynamics and in particular bifurcations of the important internal periodic orbits within the NHIMs. These changes
also include the creation of chaos in the internal dynamics. However, according to the persistence theorem, the NHIM survives and keeps its topology as long as the normal hyperbolicity property
is conserved. The development scenario of the internal dynamics of a NHIM under any perturbation
can be investigated and presented graphically best by the restriction of the Poincar\'e map to the
NHIM.

In the following, we are interested in Hamiltonian systems with 3 degrees of freedom (3-dof).
They have a complete phase space of dimension 6, a constant energy manifold of dimension 5,
and Poincar\'e maps acting on domains of dimension 4. Therefore, the most interesting invariant subsets in
the constant energy manifold are the ones of dimension 3, and they correspond to invariant surfaces of
dimension 2 in the domain of the Poincar\'e map. In most cases, it is simpler to argue for maps
than for flows. Therefore in this article, we will present our ideas mainly for maps.
For 3-dof systems, the codimension-2 NHIMs in the map are 2-dimensional and then the map
of their internal dynamics can be displayed by 2-dimensional graphics. We use this graphical representation
extensively in the present article.

Some investigations of the loss of normal hyperbolicity of NHIMs have been reported in
\cite{li,inerrea,allahem,mauguiere}.  These examples either use 2-dof systems or use
simplifications which are equivalent to a reduction of the problem to 2 degrees of freedom. They are mainly
concerned with the implications of the loss of normal hyperbolicity to transition state theory and
transport problems. They are less concerned with the remnants of the NHIMs
during their decay to fragments which are partially of lower dimensions, a problem with which we
are concerned in the present article.  For further interesting publications on
changes of NHIMs and their invariant manifolds under parameter changes see also \cite{teramoto,mackay,teramoto1,teramoto2,ketzmerick0,ketzmerick1,main,llave}.

In the present article, we deal with the problem of the disintegration of NHIMs when they lose their
normal hyperbolicity and with the problem of how we observe this decay.  We investigate the
remaining fragments numerically with the help of phase space structure indicator functions.
Then let us give here some basic introductory remarks on indicator functions.

When a 0-dimensional unstable invariant set, i.e. a hyperbolic fixed point of a map, collides with
a stable fixed point and runs through a saddle-center bifurcation, then it disappears completely
from the domain of the map.  Using the normal form approximation of the map around the hyperbolic
equilibrium point, we can observe how these fixed points disappear into the complex generalization of the
domain of the map.

First, we consider a set of parameter values where the unstable fixed point still exists. Let us
discuss what we observe in indicator functions. The most appropriate indicator function for this
purpose is the time delay function which indicates the time during which a general trajectory stays in some
neighbourhood of a particular region, for example, an unstable fixed point. 
If the trajectory starts exactly on the stable manifold of the fixed point, then this trajectory converges to the fixed point, it stays in the
neighbourhood of the fixed point for an infinite future and accordingly, the time delay diverges.
When it starts not exactly on the stable manifold but close to it, then this trajectory comes
close to the fixed point, stays in its neighbourhood for some time but leaves this neighbourhood
again after a finite time and accordingly, the time delay is large but finite.

If we plot this time delay over some surface of initial conditions in the phase space, then we detect
the intersections between the stable manifold of the fixed point and this surface. This basic idea also
works for the detection of stable manifolds of higher-dimensional unstable invariant subsets,  
and it is the basic idea of how we search for NHIMs numerically. For all the details, see \cite{gonzalez}.

For a 0-dimensional unstable fixed point, the invariant set disappears at once completely in a
single bifurcation at one specific value of the perturbation parameter. For higher dimensional
invariant subsets, the decay of the invariant set is a process which starts at a particular value
of the perturbation parameter. Then the decay continues over some possibly large
interval of the perturbation parameter. Parts of the original 2-dimensional NHIM surface remain
while other parts decay and leave behind only a ( in general fractal ) collection of fragments which
partially have lower dimensions than the original NHIM. In the most extreme case, the NHIM
in the map turns into a fractal powder only; for an example of such a scenario see \cite{gonzalez2}.

In the time delay function, we recognise these events as follows, when we construct the indicator
function on a 2-dimensional domain, i.e. on a surface of codimension 2 in the domain of the map.
The stable manifold of the NHIM in the map is a fractally folded submanifold of dimension 3 before
the beginning of the decay. So its transverse intersection with a codimension 2 surface is a fractal
collection of 1-dimensional curves. Surviving parts of the NHIM also give these results after
the start of the decay.

In contrast, look at decaying parts. Here the NHIM surface decays into lower than 2-dimensional
parts. Let us look at a particular part with a ( maybe fractal ) dimension $k$. Their stable
manifolds have dimension $k+1$. The transverse intersection of these stable manifolds with the
2-dimensional intersection surface over which we plot the indicator function has dimension $k-1$.
In particular, if $k<1$ then this intersection is empty in general. That is, we do not see
any corresponding singularities in the plot.

However, when we start in regions close to some remnants of the stable manifolds of the NHIM then
there is some large but finite time delay compared with other regions nearby. The effect looks as if the
infinite singularities of the indicator function are removed and only a high background is left. If there is a
fractal collection of remnants, then the time delay function has a lot of finite maxima which gives this
function a complicated appearance. For a previous observation of this type of complicated behaviour
in a time delay function along a 1-dimensional line of initial conditions, see \cite{jung}.

These results form a new type of transient behaviour which shows an interesting contrast to the
usual transient behaviour near unstable invariant subsets. For a NHIM before the beginning of
any decay, we are used to the following behaviour. A general trajectory comes close to the
invariant subset moving near the stable manifold of this subset. Then it stays near the
invariant subset for some finite time and leaves the neighbourhood of this subset again moving
close to the unstable manifold of this unstable invariant subset. While close to the invariant
subset the general trajectory performs a type of motion very similar to the motion of
trajectories within the invariant set. This is the usual type of transient behaviour, see
\cite{tel}. So we have transient dynamics in the normal direction of the NHIM and this usual
transient motion continues also during the decay of the NHIM.

However, when the NHIM starts to decay, we find in addition also some kind of transient
behaviour in the tangential direction of the NHIM. This can be understood along the following
considerations. Usually, a NHIM which did not yet start its decay is compact, and then a trajectory
belonging to the NHIM can not leave the NHIM. In addition, a trajectory very close to the NHIM can
not leave the neighbourhood of the NHIM in the tangential direction. It will do so only in the normal
direction.

This situation changes fundamentally as soon as the NHIM starts its decay. Then
a general trajectory belonging to the NHIM and not moving on a regular substructure
can approach the boundary of decay by tangential motion on the NHIM. It can finally
cross this boundary also by tangential motion along the surface which has belonged to the
NHIM before the beginning of the decay. After the beginning of the decay, such regions are
the logical continuation of the surviving parts of the NHIM in the tangential direction.
Thereby this general trajectory is lost from the neighbourhood of the remaining pieces of the NHIM.

Finally, we consider a general trajectory that approaches the neighbourhood of the
NHIM by motion close to its surviving parts of the stable manifold. Then this trajectory in
the neighbourhood of the decaying NHIM has two competing possibilities to leave the
neighbourhood of the surviving parts of the NHIM again. First, it can leave along the unstable
manifolds of the surviving parts of the NHIM. Second, it can move mainly tangentially to the NHIM
over the boundary of decay and leave the surviving parts of the NHIM along this direction.
Or more generally, it can leave by a combination of these two routes.

In this article, we treat the case of 3-dof systems and NHIMs of dimension 2 in the map.
When such a NHIM starts to decay, then we find surviving parts of 2-dimensional surfaces.
In the general case, the internal dynamics of such parts contain KAM curves and chaos
layers. Then the interior of KAM curves ( and this includes chaos layers inside ) are trapped
on these surviving parts of the NHIM forever.

As a short side remark let us mention briefly the essential differences which we find for systems
with even more degrees of freedom, let us say $k>3$ in number. Then the dimension of the domain
of the map is $2k-2$. Now assume that we have a codimension 2 NHIM, i.e. it has dimension
$2k-4>2$ in the map. There are still KAM curves of the internal dynamics of the NHIM. And we
have chaos layers. However, for $k>3$, we have Arnold diffusion in the internal dynamics of the
NHIM, all the chaos layers are interconnected and a general trajectory starting in some fine chaos
layer in the long run can diffuse over the whole chaotic part of the internal dynamics.

Now assume that the decay of the NHIM has started in some part. The boundary of the decay
has contact with this global chaotic part. And accordingly, all the trajectories starting in
the chaotic part will cross the boundary of the decay in the long run. In this sense, there
are no surviving parts of the NHIM of the full dimension $2k-4$ as soon as the decay has
started in some part of the NHIM. All the remnants of the NHIM are a collection of lower
dimensional subsets only. This is qualitatively different from the case of NHIMs of dimension 2,
where we find surviving parts of the full original dimension.

In this work, we take the motion of a charged particle in a perturbed magnetic dipole field
as an example of demonstration. The perturbation is the addition of a quadrupole contribution
to the magnetic field. This example has one unusual property. Namely, the decay of the NHIM
starts as soon as the perturbation parameter, i.e. the magnitude of the quadrupole contribution
becomes different from 0. So we do not have some interval of the perturbation parameter, where
the NHIM persists in the topology which it had for perturbation 0. However, this is no
problem for the topic of the present article, since we are interested in the decay of NHIMs anyway.
Otherwise, this magnetic dipole example is an excellent and typical example to study the decay
scenario of a NHIM. Details of our previous studies of this magnetic dipole example and many
further references for this system can be found in \cite{gonzalez1}.

The present article is organised as follows:
In section 2, we present the basic dynamics of our example of demonstration,
in particular, the construction of the main NHIM and its stable and unstable manifolds.
In section 3, we give a short explanation of our method to search for and represent the Poincar\'e
map of the internal dynamics of the NHIMs. In section 4, we show the plots of the
indicator function, i.e. the time delay function, and for comparison also the
corresponding plots of the Poincar\'e map restricted to the NHIM. The last section
contains our conclusions and final remarks.

\section{Example: The motion of an electrically charged particle around a perturbed  magnetic dipole }

As an example of demonstration for the disintegration of a NHIM and
for the creation of corresponding tangential transient effects, we treat
the motion of an electrically charged particle in the field of a
magnetic dipole perturbed by a magnetic quadrupole. The Hamiltonian function of the system in
cylindrical coordinates $(r,\phi,z)$ is given by

\begin{equation}
    H =  \frac{1}{2m} \left(\vec{p} - \frac{q}{c} \vec{A} \right)^2 = \frac{1}{2m} \left( p^2_{r}  + \left( \frac{p_{\phi}}{r} - A_0 \frac{r}{ (r^2 + z^2)^{3/2} } \right)^2 + \left(  p_z - \epsilon \; \frac{r^2 \sin{2\phi} }{ \left( r^2 + z^2 \right)^{5/2} } \right)^2  \right).
    \label{eq1}
\end{equation}

Here $A_0$ is the magnitude of the magnetic dipole moment which is aligned
in the $z$-direction. The term in Eq.\ref{eq1} containing $A_0$ is the magnetic dipole
potential. And $\epsilon$ is the magnitude of the quadrupole perturbation of
the field, which serves as a perturbation parameter in the investigation of
the development scenario of the NHIM of this particular system.
The term in Eq.\ref{eq1} containing  $\epsilon$ is the magnetic potential of one particular
quadrupole component. The phase space of the full system is 6-dimensional and has the
coordinates $(r, p_r, \phi, p_\phi, z, p_z )$.  Let us consider a fixed value of the total energy $E$
for the next considerations and denote as $M_0$ the NHIM of the system for the
unperturbed case and $M_\epsilon$ the NHIM for the perturbed case.

The dipole potential has a rotational symmetry around the $z$-axis and
correspondingly under the influence of the dipole field only, i.e.
in the unperturbed case $\epsilon=0$ the $z$-component of the angular momentum
of the particle, i.e. $L_z = p_\phi$, is conserved and it acts as a parameter of the system. 
In this case, the total 3-dof system reduces to a 2-dof system with 
an effective potential energy is given by

\begin{equation}
    V_{eff}(\rho,z)  = \frac{1}{2m}  \left( \frac{L_z }{r} - A_0 \frac{r}{ (r^2 + z^2)^{3/2} } \right)^2  .
\end{equation}

This 2-dimensional effective potential energy has a saddle point which generates a  Lyapunov family of unstable hyperbolic periodic orbits $\gamma_{L_z}$ in the phase space of the 2-dof system 
when we change the value of $L_z$ \cite{gonzalez, gonzalez1}.  Those hyperbolic periodic orbits parametrized by the $L_z$ are the basic
constructing blocks of the NHIM in the phase space of the 3-dof symmetric unperturbed system. To construct
the NHIM $M_0$ for this system, let us consider the union of all the hyperbolic periodic orbits in the Lyapunov family and take
the Cartesian product with the circle $S^1$ that represents the angle $\phi$. Then the NHIM $M_0$ of the 3-dof
unperturbed symmetric system is

\begin{equation}
M_0 = \bigcup_{L_z} \gamma_{L_z} \times S^1.
\end{equation}
In the same way, we construct the stable and unstable manifolds  $W^{s/u}(M_0)$ of the NHIM $M_0$ 
taking the union of all the stable and unstable manifolds  $W^{s/u}(\gamma_{L_z})$  of the hyperbolic periodic orbits $\gamma_{L_z}$ and 
the Cartesian product with the circle $S^1$,

\begin{equation}
W^{s/u}(M_0) = \bigcup_{L_z} W^{s/u} (\gamma_{L_z}) \times S^1.
\end{equation}  
For the quadrupole field, we have chosen a component which is not symmetric around the
$z$-axis such that for $\epsilon \ne 0$ the angular momentum $L_z$ is no
longer conserved. In this case, the system can no longer be reduced to
a 2-dof system.

In the next section, we show numerical results for the perturbation of the
system where we start with the unperturbed case of $\epsilon = 0$ and then
increase $\epsilon$ slowly to observe the effects of the perturbation away
from the partially integrable case. 

In particular, we will study the change of the NHIM $M_\epsilon$
of the system under this perturbation. Therefore, it is important to understand the
structure of the NHIM $M_0$ well in the unperturbed case. For $\epsilon=0$, the 6-dimensional phase
space of the complete 3-dof system foliates into a continuum of 2-dof
systems, one for each possible value of the conserved quantity $L_z$.
The phase space of each one of these reduced 2-dof systems is a copy of
the 4-dimensional space of the variables $(r, p_r, z, p_z)$.

Accordingly, the full 6-dimensional phase space is a Cartesian product of
a circle representing the cyclic angle $\phi$ with the pile of the 4-dimensional
phase spaces of the reduced 2-dof systems. During this pile construction
the pile parameter $L_z$ takes over its role as a phase space coordinate
of the full 3-dof system.

Having this pile construction in mind it is easy to imagine the structure of the
NHIM $M_0$ of the unperturbed full 3-dof system. First, let us consider one particular
value of the total energy $E$. This cuts out a 5-dimensional manifold of the
full phase space and a corresponding pile of 3-dimensional energy manifold
of the reduced 2-dof systems.

This pile construction also induces a pile construction of the Poincar\'e map
and a foliation of its 4-dimensional domain into 2-dimensional leaves
belonging to the various values of $L_z$. Now imagine that the reduced
Poincar\'e map has a hyperbolic fixed point $X_{L_z}$ which exists for some interval
of $L_z$ values. In the pile construction, this continuum of fixed points
forms a continuous line of points and by the formation of the Cartesian product with the
circle $S^{1}$ representing the cyclic angle $\phi$ it forms a 2-dimensional manifold with the topology
of a cylinder segment.  We denote the NHIM in the Poincar\'e map as $M^P_0$.
Then, the NHIM $M^P_0$ is given by

\begin{equation}
M^P_0 = \bigcup_{L_z} X_{L_z} \times S^1.
\end{equation}

This manifold is invariant under the Poincaré map by construction, and it inherits
normal hyperbolicity from the hyperbolicity of the fixed point in the reduced
system. Thereby this surface is the NHIM $M^P_0$ of the unperturbed partially
integrable system in the 4-dimensional domain of its Poincar\'e map. 
In an analogous way, we can construct the stable and unstable 
manifolds of the NHIM $M^P_0$ corresponding to the Poincaré map as

\begin{equation}
W^{s/u}(M^P_0) = \bigcup_{L_z} W^{s/u} (X_{L_z}) \times S^1.
\end{equation}   

In this particular example,  the normal hyperbolicity
of the unperturbed NHIM $M^P_0$ vanishes at one boundary of the cylinder segment and thereby it is
clear that any perturbation of the partial integrability of the system triggers the decay
of this NHIM $M^P_0$ surface beginning from this boundary with vanishing normal
hyperbolicity. Therefore, this dipole example is an ideal system for studying the
events occurring during the decay of a NHIM $M_\epsilon$ under well-controlled conditions.
For more details on all these considerations, including detailed numerical
illustrations, see  \cite{gonzalez,gonzalez1} and references therein.

\newpage

\section{The Numerical Algorithm to Follow a NHIM
Along a Curve of the Perturbation Parameter}

We give a short description of our method to search for NHIMs and for the pictorial presentation of their internal dynamics. And we include comments which may help to clarify our method.
  
 The method begins with a search for the
 local branch of the stable manifold of an unstable invariant subset of codimension
 2 in the Poincar\'e map. This stable manifold has codimension 1 and therefore we
 expect that a curve of dimension 1 placed into an appropriate region
 of the domain of the map intersects this stable manifold in isolated points. We
 search for one of such intersection points.
 
 For the starting case, $\epsilon = 0$, we know the NHIM and its position. Let us call it $M_0$.
 And now we search for the NHIM for some small value of $\epsilon \ne 0$. Let us call it 
 $M_{\epsilon}$. Because of continuity and persistence, $M_{\epsilon}$ is located near $M_0$. So we
 select some tubular neighbourhood $U$ around $M_0$ such that we are confident that $M_{\epsilon}$
 lies within $U$. Next, we place some curve $C$ into $U$ such that we suppose that
 $C$ intersects the local branch of $W^u (M_{\epsilon}) $. We place many initial points
 along $C$ and calculate for each one of these points the number of iterations of the
 full 4-dimensional Poincar\'e map which lie still within $U$.
 
 If one of these points would lie exactly on the local branch of the stable manifold of an invariant subset,
 then this number of iterations within $U$ would be infinite. Of course, numerically
 we only detect the region of $C$ where this number becomes maximal. Then, we zoom in as
 fine as our numerical precision allows to get this maximum with the best possible precision.
 The corresponding trajectory ( sequence of iterated points ) first approaches the
 invariant subset along its stable manifold, stays close to the subset for a while
 and finally leaves the neighbourhood of the invariant subset again along its unstable
 manifold. Now, we truncate this trajectory sufficiently long before it leaves $U$
 such that we are confident that the last point we keep is still very close to the
 invariant subset.
 
 We use this last point which we keep as our new initial point. We place an appropriate
 new curve $C$ through this last point. This time $C$ can be rather short. And we
 search for the point along the new curve $C$ which keeps the trajectory inside of
 $U$ for the longest possible sequence of iterated points. And we continue this
 procedure. Thereby we continue the sequence of points in the close neighbourhood of
 $M_{\epsilon}$ until we have a long sequence of points very close to the NHIM we are
 looking for.
 
 So the procedure consists of iterated applications of the full 4-dimensional map with repeated
 interruptions to perform the fine-tuning along some curve $C$. Many times it is convenient
 to place the curve $C$ into the direction of the momentum normal to the NHIM surface.
 This fine-tuning is a version of the control of chaos where we keep some numerical
 trajectory on the stable manifold of some invariant subset, see also \cite{ml1, ml2}
 
 We must understand why this trajectory represents some substructure of the
 internal dynamics of the NHIM. The answer is given by the foliation theorem of stable
 and unstable manifolds of NHIMs, see chapter 5 in \cite{wiggins}. Basically ( expressed in a
 somewhat oversimplified manner ) it says that these
 stable and unstable manifolds are the Cartesian product of a line with the internal
 structure of the NHIM. Then our initial point belongs to some internal structure
 of the NHIM and the whole trajectory stays on this particular internal structure
 and represents it. Behind this foliation property stands the property of the full
 dimensional Poincar\'e map in the neighbourhood of the NHIM to have a natural
 factorization into a normal factor and a tangential factor. Our numerical
 procedure filters out automatically the tangential factor.
 
 For some well-chosen collection of initial curves $C$ placed over the various
 important internal structures of the NHIM, we can produce a corresponding collection
 of trajectories on the NHIM which represent all important internal structures of
 the NHIM. To this end, we can try out many initial curves $C$ and select the
 ones which lead to a good pictorial presentation of the NHIM structures, including
 important KAM islands, important secondary structures, chaos stripes, etc, in
 analogy to what we are used to do when constructing plots of 2-dimensional Poincar\'e maps.
 By this procedure, we obtain simultaneously the position of the NHIM and a pictorial
 representation of its internal dynamics. The plots we show are the projection of
 the 2-dimensional NHIM surface into some coordinate plane of the domain of the
 4-dimensional Poincar\'e map. Preferably, we use a coordinate plane giving a
 1:1 projection, whenever this is possible.
 
 Now assume that we have the NHIM for some value $\epsilon_1$ of the perturbation
 and want to construct it for
 some new value $\epsilon_2$ which is a little larger. Then we place a new neighbourhood
 $U$ around the $M_{\epsilon_1}$ and apply the same idea as before to construct
 the NHIM for the new larger perturbation. So we proceed in sufficiently small steps
 towards larger perturbation. At some point, we come to a value of $\epsilon$ where
 the NHIM starts to lose normal hyperbolicity at least locally at some point.
 
 Here the reader may ask how we notice the loss of normal hyperbolicity and how we
 can check normal hyperbolicity. First, the whole search strategy relies heavily
 on the existence of a stable manifold of codimension 1, and this, in turn, relies on
 the normal hyperbolicity. In this sense, we can interpret the success of our method
 as a numerical indication of the presence of normal hyperbolicity at least in the local region of
 the NHIM where we are looking for some internal NHIM structure. There are 
 many examples of the success of this approach, see e.g  Fig.7 in \cite{gonzalez2},
 Figs.7 and 8 in \cite{jz}, Figs.12 and 13 in \cite{zj1}, Figs.16, 17, 18, and 19
 in \cite{zj2}, Figs.14 and 15 in \cite{zj3} and Fig.13 in \cite{jw} and several more
 examples which are not yet published.
 
 We can do even more to check normal hyperbolicity directly in some regions of the NHIM.
 If we know some periodic orbit within the NHIM region, then we can calculate the monodromy
 matrix of this orbit and obtain its stability properties in the tangential and in the
 normal direction of the NHIM. Then we see at which value of some parameter the
 tangential instability of the corresponding fixed point in the reduced map becomes
 larger than the normal instability, i.e. at which value it loses normal hyperbolicity.
 A nice example of this check has been shown in \cite{gonzalez2}. Here at some parameter value, an
 important fixed point on the NHIM loses normal hyperbolicity. Exactly at this
 parameter value the NHIM surface
 grows a singularity and breaks and the numerical search method starts to fail in the
 neighbourhood of this fixed point, the NHIM surface grows a hole. In \cite{jw} another
 example of the creation of a hole in the NHIM and the corresponding breakdown
 of our search strategy has been presented. Again it happens exactly at the parameter
 value at which an important periodic orbit inside the NHIM loses its normal hyperbolicity.
 The formation of the hole in the NHIM surface starts exactly at the position of
 the fixed point which loses normal hyperbolicity, compare Figs.6 and 13 in \cite{jw}.
 
 Now it is time to make a few clarifying remarks on our search strategy. It is based only
 on the very elementary qualitative properties of NHIMs and the trajectories in their neighbourhoods.
 It only needs the numerical construction of the full 4-dimensional Poincar\'e map of the dynamics or
 equivalently it only needs the numerical construction of trajectories in the full 6-dimensional phase space of the system. It never needs any reduction of the dynamics to some centre manifold or some other lower dimensional subsets nor does it need any construction of normal forms or any other analytical approximations.
 The above mentioned factorization of the dynamics
 does the job all by itself. The reduction of the dynamics to the centre manifold is
 an alternative strategy which does this factorization by analytical approximations
 and by the construction of normal forms, see \cite{wsw,wwju}.
 
 As mentioned,  this simple method is based on qualitative properties of NHIMs
 and the trajectories in their neighbourhoods only. Therefore, the main mathematical condition which
 ever enters the whole procedure is the maximization of the time that the trajectory spends in the 
 neighbourhood of the NHIM, as we change the initial conditions.
 The whole procedure is numerical and it only relies on the general qualitative properties of NHIMs.
 Of course, this has the disadvantage that we can not derive the conclusions and
 explanations of the dynamics which one might draw from reductions and normal forms
 \cite{jm,gmm,ss} or the parametrization method \cite{haro}. 
 However, our method can give us important details about the rich bifurcation process 
 regardless of any hypothesis considered in analytical methods.

 Finally, we proceed to some remarks on the events which happen as soon as
 the decay of the NHIM has started. Suppose that a chaos stripe comes into contact with
 the decay boundary of the NHIM and we start some trajectory over the part of this
 chaos strip which still lies behind the boundary in the part of the NHIM which still
 shows normal hyperbolicity. Then we obtain an initial part of the trajectory
 belonging to the chosen initial point by our search method. However, after some
 number of iterations, this trajectory crosses the decay boundary and enters a region
 where normal hyperbolicity is already lost. Then also the stable manifold as a
 codimension 1 surface is lost. In the best case a collection of lower-dimensional  
 fragments still exist.
 
 Then our search method no longer finds a clearly peaked maximum of the number
 of iterations along the curve $C$. It only finds a region of high values with
 several relative maxima and minima, as a typical example see Fig.4 in \cite{jung}.
 Then the intent to zoom in on some true infinity fails and the method breaks down.
 We interpret the initial segment of the trajectory up to the breakdown of the
 search method as a transient trajectory on the NHIM. In our example of the
 perturbed magnetic dipole such chaos regions with contact with the decay boundary
 can have large areas and accordingly, they have a large impact on plots of indicator
 functions. This is the topic of the present article.

\section{Decay of the NHIM: Numerical Results for the Delay Time and for the Poincar\'e Map }

The delay time is a natural tool to study the phase space of open Hamiltonian systems.
In general, the phase space structure indicators are useful to visualize invariant manifolds
in the phase space. In particular, they are useful to find KAM islands, unstable and stable
manifolds of NHIMs and to obtain information on their bifurcations when we change the parameters
in the system. The details of the ideas of phase space structure indicators like the fast
Lyapunov indicator, Lagrangian descriptors, Birkhoff averages, the classical action, and delay time can be found in the references
\cite{gonzalez1,lega,wiggins5,mancho,lopesinos,meiss,gonzalez3}.
For more information on the recent developments in this topic, see the references contained in
\cite{agauglou,skokos}.

The delay time $t_d$ as a phase space structure indicator is defined as the sum of two parts,
the time delay $t^+_d$ calculated forwards and the time delay backwards $t^-_d$.

\begin{equation}
t_d = t^+_d + t^-_d = \tau - \frac{r^+-r_0}{v^+} + \tau - \frac{r^- - r_0}{v^-}. 
\end{equation}

Here $r_0$ is the distance of the initial point to the origin at the initial time, $r^+$ the distance at the time
$\tau$ of the forward integration, $r^-$ the distance at time $\tau$ of the backward integration, and $v^+$
and $v^-$ are the asymptotic velocities of the particle. For this Hamiltonian system, both velocities coincide
$v^+=v^-$. More details about the delay time as a phase space structure indicator are contained in the reference
\cite{gonzalez1}.

To visualize the decay of the NHIM $M_0$ and the transient behaviour of the trajectories, we break the
rotational symmetry and calculate the time delay indicator on a surface of initial conditions that intersects
the perturbed NHIM $M_\epsilon$ and its stable manifold. This phase space structure indicator is a
natural choice for open systems. It gives us direct information on the trajectory's behaviour.
If the trajectories are close to the bounded invariant chaotic set, then their time delay is larger
compared to the other trajectories not approaching the invariant set. Only the trajectories on the
fractal chaotic set generated by the stable and unstable manifolds of the NHIM $M_\epsilon$ and the ones in the stable
KAM tori stay trapped in the interaction region forever.

\begin{figure}[H]
    \centering
    \includegraphics[width=0.7\linewidth]{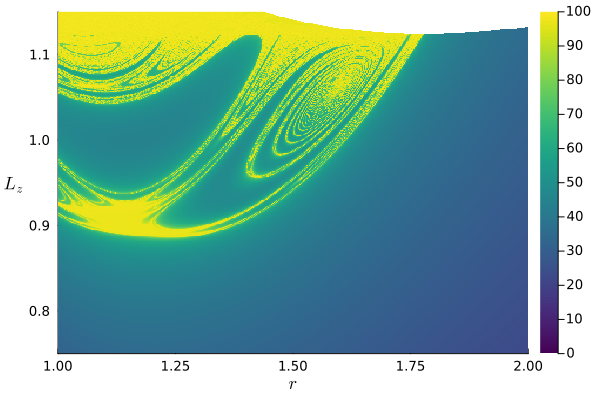}
    \caption{Time delay indicator on a colour scale for the unperturbed case, $\epsilon=0$ ( see the homoclinic tangle formed by the stable and unstable manifolds of the NHIM $M_0$ in Fig.1 in \cite{gonzalez0} and Fig.2 in \cite{gonzalez} for different values of $L_z$ ). The integration time $\tau=50$. }
    \label{fig:1}
\end{figure}

First, we consider the unperturbed system as a reference for studying the perturbed case.
We take initial conditions on the $r-L_z$ plane at $z=0$, $p_r = 0$,
and $\phi=0$ for the fixed value of the energy $E=0.05$, and calculate the time delay. The results are shown
in the Fig.\ref{fig:1} on the colour scale. The blue regions correspond to low escape time,
the trajectories with those initial conditions escape to the asymptotic region faster than the other
regions. The yellow-green fractal corresponds to the intersection of the homoclinic tangle of the NHIM $M_0$
and its stable manifolds with the set of initial conditions. The yellow region with low values of $L_z$ are the intersections with
the stable manifolds of the KAM islands, those trajectories on the invariant sets are trapped forever. The right border of
this fractal corresponds to the NHIM $M_0$. For more information on the bifurcation diagram of the unstable
periodic orbits and stable periodic orbits of the system see Fig.\ref{fig:1} (b) in reference \cite{gonzalez1}.

  To give an interpretation of the figures of the indicator function, it is
  instructive to compare them with the corresponding plots of the inner NHIM $M_0$
  structure. These plots are constructed by the following method: First, we
  construct the Poincar\'e map in the intersection surface $z=0$ where the
  intersection orientation is irrelevant because of the reflection symmetry
  of the system in this intersection plane. Second, we restrict this map to
  the 2-dimensional NHIM $M^P_0$  surface. Thereby, we obtain the Poincar\'e map for
  the internal dynamics of the NHIM $M_0$, which is a 2-dof dynamics corresponding
  to a Poincar\'e map having a 2-dimensional domain. In the following, we call
  this map of the internal dynamics of the NHIM ``the restricted map''.

\begin{figure}[H]
    \centering
    \includegraphics[width=0.7\linewidth]{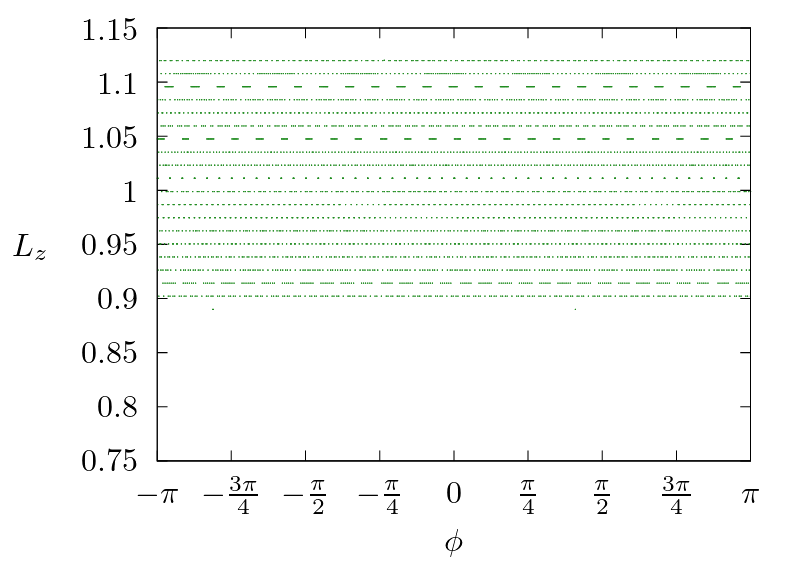}
    \caption{Poincar\'e map of the NHIM $M_0$ for the unperturbed case $\epsilon=0$.}
    \label{fig:5}
\end{figure}

\begin{figure}[H]
    \centering
    \begin{tabular}{c c c}

      \subf{\includegraphics[scale=0.275]{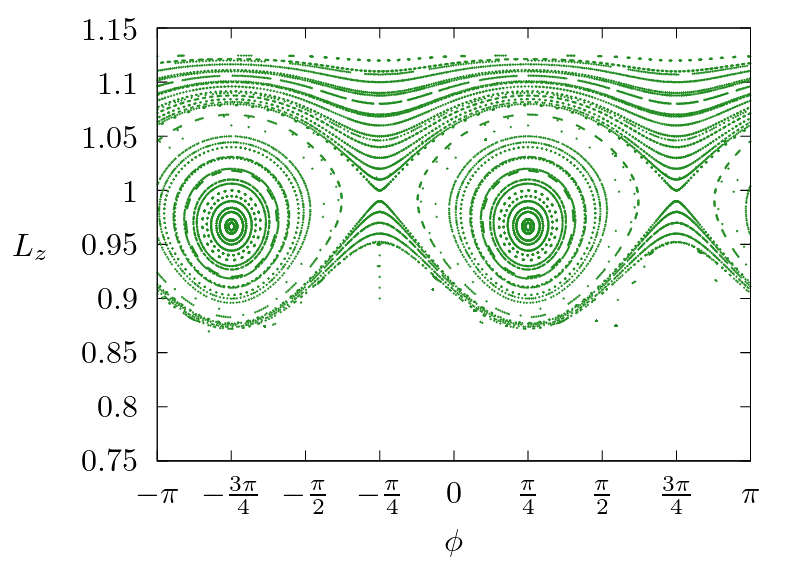}}
     {(a) $\epsilon = 0.05$ }
     &
     \subf{\includegraphics[scale=0.275]{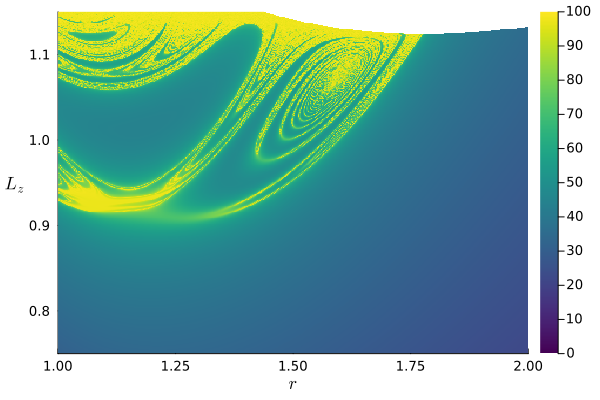}}
     {(b) $\epsilon = 0.05,\; \phi = -\pi/4$ }
     &
     \subf{\includegraphics[scale=0.275]{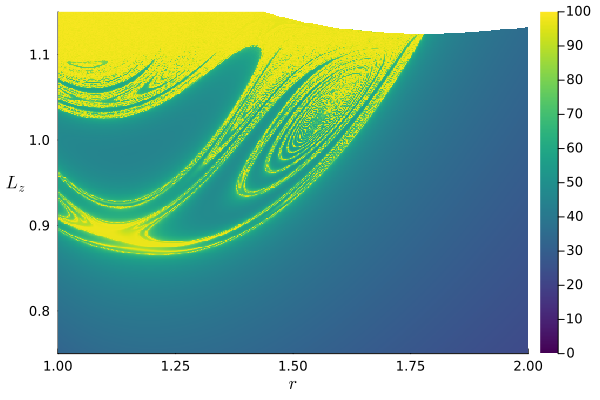}}
     {(c) $\epsilon = 0.05,\; \phi = \pi/4$ }
     \\
     \subf{\includegraphics[scale=0.275]{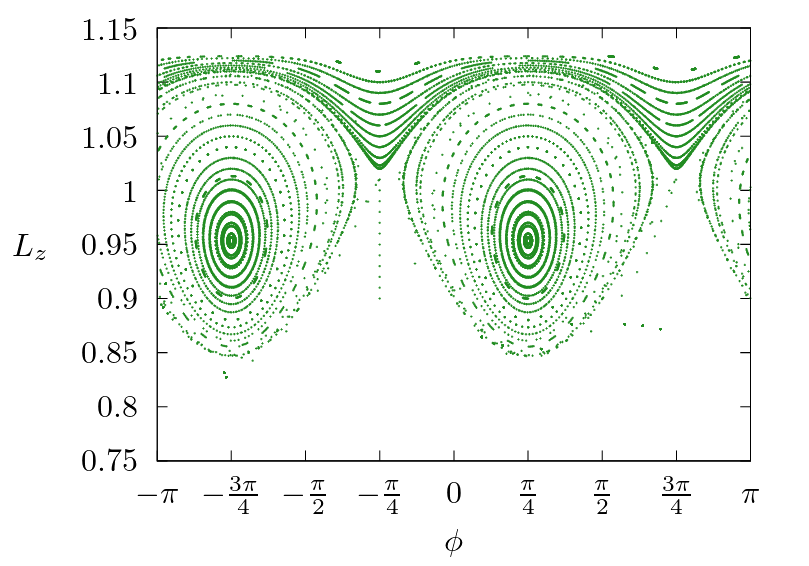}}
     {(d) $\epsilon = 0.1$ }
     &
     \subf{\includegraphics[scale=0.275]{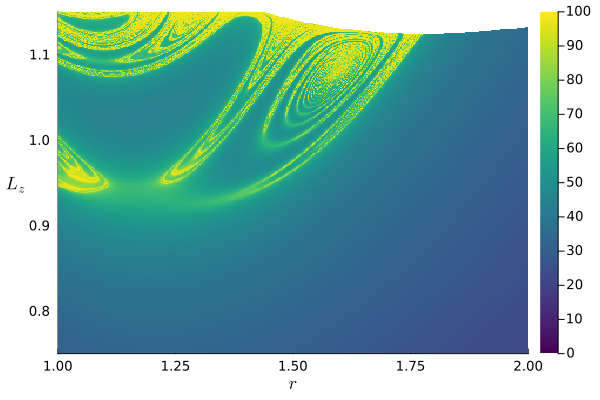}}
     {(e) $\epsilon = 0.1,\; \phi = -\pi/4$ }
     &
     \subf{\includegraphics[scale=0.275]{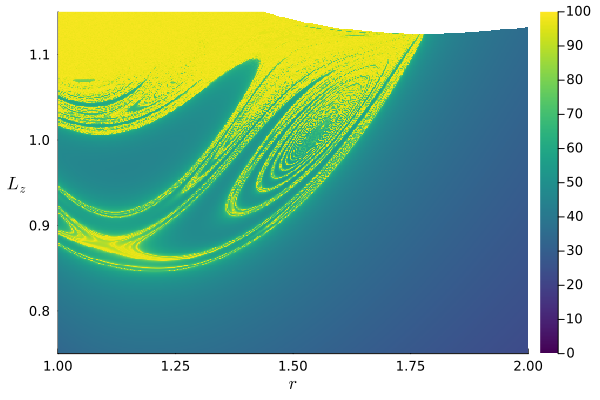}}
     {(f) $\epsilon = 0.1,\; \phi = \pi/4$ }
     \\
          \subf{\includegraphics[scale=0.275]{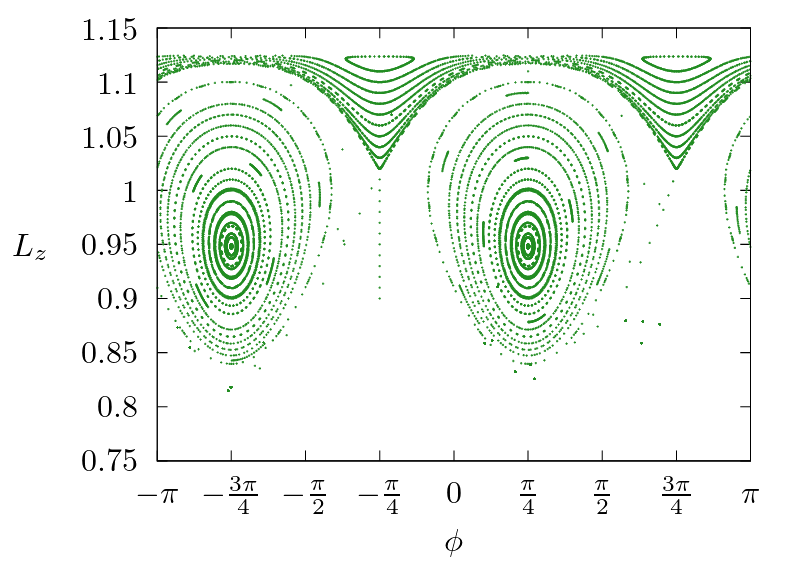}}
     {(g) $\epsilon = 0.12$ }
     &
     \subf{\includegraphics[scale=0.275]{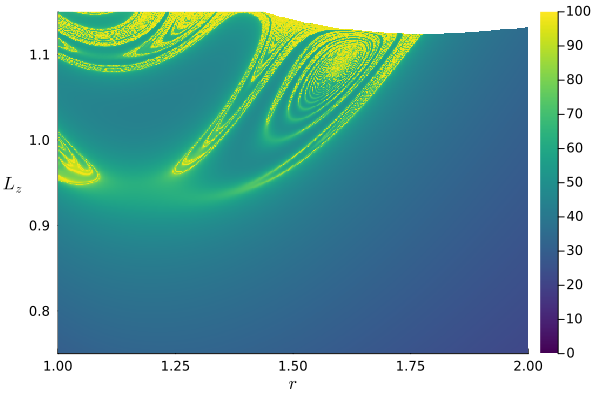}}
     {(h) $\epsilon = 0.12,\; \phi = -\pi/4$ }
     &
     \subf{\includegraphics[scale=0.275]{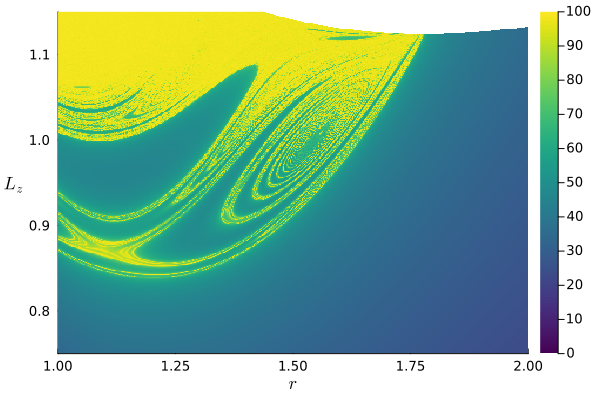}}
     {(i)  $\epsilon = 0.12,\; \phi = \pi/4$}
     \\
     \subf{\includegraphics[scale=0.275]{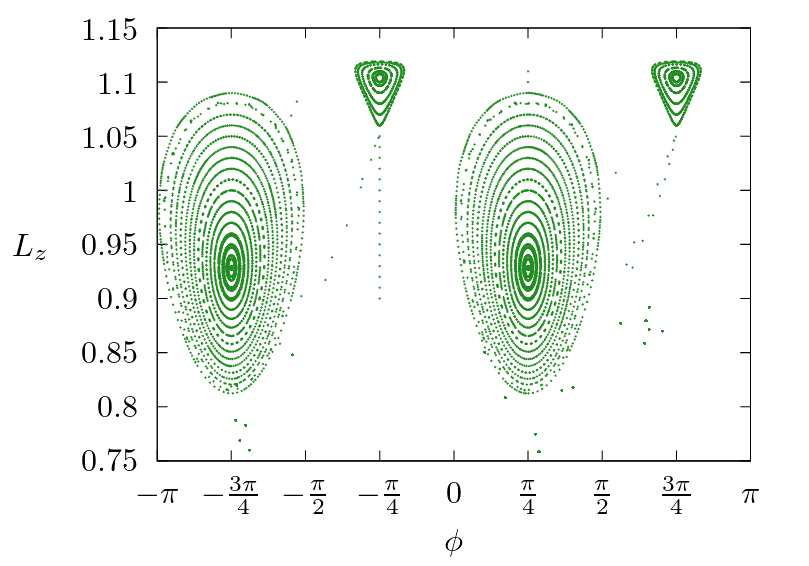}}
     {(j) $\epsilon = 0.2$ }
     &
     \subf{\includegraphics[scale=0.275]{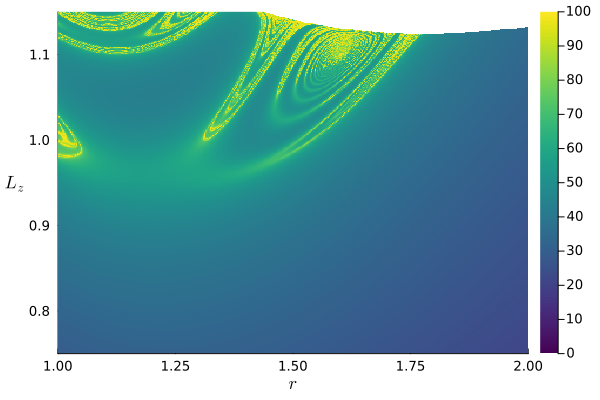}}
     {(k) $\epsilon = 0.2,\; \phi = -\pi/4$ }
     &
     \subf{\includegraphics[scale=0.275]{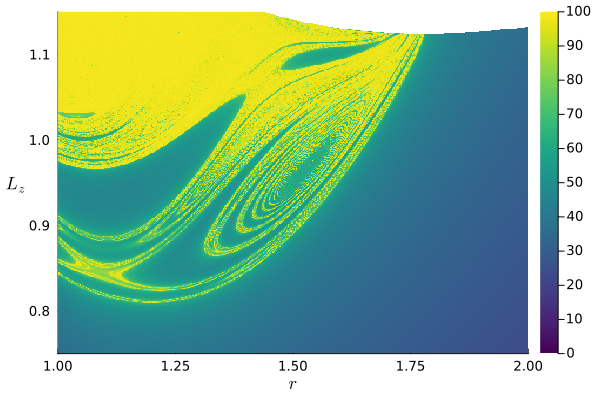}}
     {(l)  $\epsilon = 0.2,\; \phi = \pi/4$}
    \end{tabular}
    
    \caption{ On the left column, the Poincar\'e maps of the NHIM $M_\epsilon$ for the perturbation parameter values $\epsilon=0.05, 0.1, 0.12, 0.2$. On the central and right column the time delay function plotted on the $r - L_z$ plane for the 4 values $\epsilon = 0.05, 0.1, 0.12,$ and $0.2$ of the perturbation parameter and for the 2 values $\phi = -\pi/4$ and $\phi = \pi/4$ of the azimuth angle. The integration time is $\tau = 50$. The value of the time delay is colour-coded according to the right-hand colour bar.
    }
    \label{fig:2}

\end{figure}

  The Poincar\'e map of the NHIM $M_0$ has a 1:1 projection on the $\phi$--$L_z$
  plane and therefore we show plots of the restricted map by this projection
  on the $\phi$--$L_z$ plane. As usual, we present the map graphically by showing
  many iterates of a moderate number of initial points, for more details see \cite{gonzalez1}. In Fig.\ref{fig:5}, we show this Poincar\'e map for the unperturbed case $\epsilon = 0$ and
  in the various parts on the left column of Fig.\ref{fig:2} we show the perturbed Poincar\'e map for  4 values of $\epsilon$.

  In Fig.\ref{fig:5} for the unperturbed case $\epsilon=0$ we see a foliation of the domain
  of the map into invariant horizontal lines of constant $L_z$. This is caused
  by the conservation of $L_z$ for the rotationally invariant case of $\epsilon=0$.
  As usual, under perturbations, the invariant lines with a rational winding number are broken and replaced
  by secondary island chains and chaos stripes whose
  width grows with increasing perturbation. The most prominent secondary structure
  is one of the islands centred at $L_z \approx 0.95$ and $\phi = \pi /4$ or
  $\phi = - 3 \pi / 4$. To this secondary structure also belongs a separatrix region
  with tangentially hyperbolic points near $ L_z \approx 1 $ and $ \phi = - \pi / 4$
  or $ \phi = 3 \pi / 4$. At the perturbation $\epsilon = 0.05$ this separatrix
  structure is still close to a separatrix line, see Fig.\ref{fig:2} (a).

  In Fig.\ref{fig:2} (a) we also observe how the decay starts from small values of $L_z$.
  However, the decaying region is still separated from the separatrix structure
  mentioned above by primary KAM curves in between. For the internal dynamics restricted to the
  NHIM $M_0$ and therefore also for the restricted Poincar\'e map, the KAM curves are impenetrable.
  Accordingly, the separatrix structure still is a truly invariant remaining part
  of the NHIM $M_\epsilon$ for such small perturbations.

  The situation changes drastically for larger values of the perturbation. In Fig.\ref{fig:2} (d)
  the KAM lines below the large secondary structure have already been broken and
  thereby the former separatrix structure has established contact with the decay
  front of the NHIM $M_\epsilon$. Then strictly speaking the complete separatrix structure is
  no longer a remaining invariant part of the NHIM $M_\epsilon$. Iterates of initial points
  from this separatrix region at first move chaotically inside of the separatrix
  structure. However after a sufficient number of iterations, they come close to the
  decay front, possibly cross it and leave the surviving NHIM $M_\epsilon$ surface tangentially.

  To illustrate this effect pictorially, several initial points on the line
  $\phi = - \pi / 4 $ have been chosen and their iterates are also included in the
  figure as long as these iterated points could be stabilized on the former NHIM $M^P_\epsilon$
  surface by our method to construct the restricted map. See the irregularly scattered
  points in the former separatrix region and below the remaining large KAM islands.
  These trajectories have a transient existence on the remaining parts of the NHIM $M^P_\epsilon$
  surface only.

  For further perturbation increases, all primary KAM curves are destroyed
  and the only remaining and truly invariant parts of the NHIM $M^P_\epsilon$ are the surviving
  stable islands. This includes very tiny ones located inside of the region of
  the transient chaos. And now we have to see how this whole scenario is transferred to
  the indicator functions and observed in the corresponding plots.

  First, we have to decide, on which 2-dimensional planes $S$ we want to construct the
  indicator function to detect well the development scenario of the NHIM $M_\epsilon$
  by the observation of time delays. According to the properties of the delay function
  already mentioned above in the introduction it must be assured that $S$ intersects
  the stable manifold $W^s(M_\epsilon)$ of the NHIM $M_\epsilon$. $W^s(M_\epsilon)$ is a surface of codimension
  1, therefore a 2-dimensional plane in general position should intersect it transversally,
  as long as the plane lies close to the NHIM. Remember that the phase space coordinates
  $z$ and $p_z$ do not appear in the map.

  This property can be assured along the following considerations. We have already mentioned that the NHIM
  surface projects 1:1 into the $\phi - L_z$ plane. And on the NHIM $M_\epsilon$ the coordinates $r$ and $p_r$
  are always close to the values $r=1$ and $p_r=0$.
  Therefore the $r - L_z$ plane at $p_r =0$ and at any value of $\phi$ should
  work well. With the left column of Fig.\ref{fig:2} in mind, 
  we have chosen the two values $\phi = \pm \pi/4$ with
  the intention that the plot at one value should be heavily influenced by the big secondary
  islands and the other plot should emphasise the separatrix structure.

  Keeping in mind that $W^s(M^P_\epsilon)$ has codimension 1 in the domain of the map and
  $S$ has codimension 2, we expect that we obtain 1-dimensional intersections with $W^s(M_\epsilon)$
  on the plane $S$. Accordingly, we expect singularities of the time delay function
  along these 1-dimensional lines of intersection. In addition, we can expect that $S$
  also has intersections with the 2-dimensional NHIM surface $M^P_\epsilon$ itself at least in isolated
  0-dimensional points.

\begin{figure}[H]
    \centering
    \includegraphics[width=0.7\linewidth]{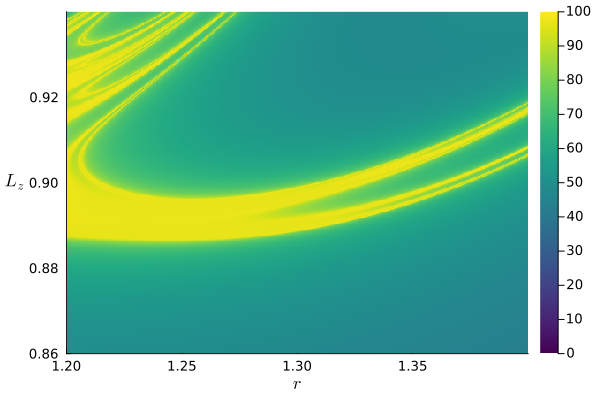}
    \caption{ Magnification of the time delay indicator from Fig.\ref{fig:1} on colour scale for the unperturbed case,
              $\epsilon=0$. The integration time $\tau=50$.}
    \label{fig:3}
\end{figure}

  Globally $W^s(M_\epsilon)$ and $W^u(M_\epsilon)$ form homoclinic intersections and build up
  a chaotic tangle. Therefore the stable and unstable manifold form a fractal of tendrils.
  In the end, we have in $S$ a fractal collection of 1-dimensional intersection lines
  with $W^s(M_\epsilon)$ leading to a fractal collection of lines of singularities of the
  time delay function. This is exactly what we see in Fig.\ref{fig:1}. Between the curves of
  singularities, the time delay drops to finite values. However, in regions of a high
  density of lines of singularities with only very small gaps in between them, also
  inside of the gaps we have rather high finite values of the time delay. In the plots,
  the whole region has obtained yellow colour. However, in a comparison between
  Figs.\ref{fig:1} and \ref{fig:3}, we see that under magnification we can resolve more gaps in 
  the fractal intersection structure.

  Next, we consider nonzero perturbations. We know already from the plots of the
  restricted map in Fig.\ref{fig:2} that along $\phi = -\pi/4$ there is a lot of decay on the NHIM $M^P_\epsilon$
  whereas along $\phi = \pi/4$ the NHIM $M^P_\epsilon$ forms a large island and thereby the NHIM $M^P_\epsilon$
  surface survives. Accordingly, on the indicator function in the plane along $\phi = \pi/4$
  sharp fractal structures remain, even though they change in some details, whereas
  in the plane along $\phi = - \pi/4$ a considerable part of the former fractal structures
  is converted into a very diffuse region of still large but not infinite values of the time
  delay.

\begin{figure}[H]
    \centering
    \includegraphics[width=0.7\linewidth]{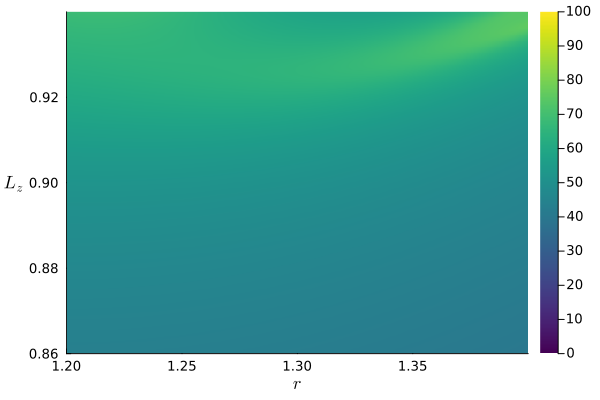}
    \caption{Magnification of time delay indicator from Fig.\ref{fig:2}  (e) at $\epsilon=0.1$, $\phi=-\pi/4$.
             The integration time $\tau=50$.}
    \label{fig:4}
\end{figure}

  In the magnification in Fig.\ref{fig:4} we observe that under a magnification no finer structure
  is resolved in the diffuse region. The delay function remains a very smooth distribution
  of high values only. There are no intersections with the stable manifold of an invariant
  surface leading to true singularities of the time delay. There is only a large region
  of initial conditions where trajectories run into the direction of the former NHIM $M_\epsilon$
  surface and stay close to this region of already destroyed parts of the NHIM $M_\epsilon$ for a large but always
  finite time.

  In the final part of this section, we point out some details of the delay plots which demonstrate how closely
  the indicator function follows the changes in the NHIM $M_\epsilon$ structure.
  In Fig.\ref{fig:2} we observe surviving fractal structures for high values of
  $L_z$ corresponding to the surviving parts of the NHIM $M^P_\epsilon$ at high values of $L_z$ also
  for $\phi$ around $-\pi/4$. This parallelism explains in which way indicator functions
  can illustrate the decay scenario of NHIMs.

  In Fig.\ref{fig:1} we observe very high values of the time delay around $\phi = 1.2$, $L_z = 0.9$
  representing a fractal of singularities which is not well resolved. Fig.\ref{fig:3} shows that
  under higher resolution and a related rescaling of the colour bar, the plot still represents a
  fractal of singularities.
  The Figs. \ref{fig:2}(b) and \ref{fig:2}(c) show that these very high values of the time delay survive the small
  perturbation of $\epsilon = 0.05$. The structures are only deformed and shifted a little.
  These structures in the delay functions are caused by initial conditions converging
  towards the separatrix structure along the part of $W^s(M_\epsilon)$ lying over the separatrix
  structure. Remember that according to the foliation theorem ( chapter 5 of \cite{wiggins} )
  $W^s(M_\epsilon)$ transports the inner structure of the NHIM $M_\epsilon$ along it.

  Next we increase the perturbation to the value $\epsilon = 0.1$ in Figs.\ref{fig:2}(e) and \ref{fig:2}(f). Here
  the very high value of the time delay for $\phi \approx 1.2$ and $L_z \approx 0.9$
  is no longer present. The better resolution
  under the magnification in Fig.\ref{fig:4} shows that for this higher perturbation, the true
  singularities pointed out above are gone. We also observe that for $\phi = - \pi/4$
  in Fig.\ref{fig:2}(e) the background value has dropped considerably whereas for $\phi=\pi/4$
  in Fig.\ref{fig:2}(f) at least some higher background can be recognised. In this form, we get a clue
  that for smaller perturbations there has been some important structure present which becomes destroyed
  for increasing perturbation.

  Finally, we point out some analogies between the indicator function and the Poincar\'e map in more detail.
  In Fig.\ref{fig:2}(a) we see that for $\epsilon = 0.05$ the large separatrix structure is still
  confined on both sides by primary KAM curves. It has not yet established any contact
  with the decay front. Therefore, trajectories in this separatrix structure cannot
  escape by tangential motion on the surviving parts of the NHIM $M_\epsilon$. This statement
  is valid for all values of $\phi$.

  The situation is drastically different for $\epsilon = 0.1$. Here the separatrix structure
  has contact with the decay front. As we have already mentioned above, initial
  conditions in the separatrix region leave by tangential motion towards the
  decay front. Accordingly, the separatrix structure is no longer an invariant part of
  the NHIM $M_\epsilon$ and also the part of $W^s(M_\epsilon)$ over the separatrix structure no
  longer exists as a true stable manifold of an invariant subset.

  By looking at Fig.\ref{fig:2}(d) it becomes understandable that initial conditions in the former
  separatrix region around $\phi=\pi/4$ and above the large secondary island need a longer
  time to diffuse tangentially to the decaying front than the ones starting around $\phi = - \pi/4$.
  Accordingly, there still exist transient remnants of the stable manifold of the former
  separatrix structure near $\phi = \pi/4$, which also have an influence on the indicator
  function when they intersect its domain, in contrast to the behaviour around $\phi = - \pi/4$.

  This explains why in Fig.\ref{fig:2}(f) we
  still observe long transients near $r=1.1, L_z=0.9$ in contrast to Fig.\ref{fig:2}(e)
  where we observe rather short transients only.
  A comparison between the whole scenarios
  of Fig.\ref{fig:2} makes it easy to understand why in total the transient effects
  fade out with increasing perturbation.

\section{Conclusions and Remarks}

The decay of the NHIM and the related tangential transient effects can be seen in
the restricted Poincar\'e map and also in the time delay function used as a phase space structure
indicator function. Now the reader may ask what are the advantages and disadvantages
of these two possibilities to display the scenario as a function of the perturbation
strength.

 The time delay function is easy and rapid to calculate and,
 it can be calculated and presented on any 2-dimensional surface in the constant energy manifold
 of the phase space. Moreover, it is not necessary to have detailed previous information on the
 location of the NHIM. The initial conditions leading to a high value of the time delay lie close
 to the stable manifold of localized subsets and guide us automatically towards the
 unstable invariant set.

  In contrast, the construction of the restricted Poincar\'e map needs the approximate
  knowledge of the position of the NHIM to start the calculations.
  As explained in detail in \cite{gonzalez} so far we use a numerical method
  which constructs the restricted map and searches for the position of the NHIM $M_\epsilon$ in a combined
  form. As long as we have small perturbations and complete NHIMs without decay this method
  works quite well and is reliable. This method can display the finest details
  of the internal NHIM dynamics in graphical form including tangentially chaotic parts.
  However, for large perturbations and NHIMs in the
  process of decay, the method becomes more difficult. In some examples ( not yet
  published so far ) it has become extremely difficult to find transient parts of the NHIM in the process of decay by this method. It needed considerable numerical effort and a special adaptation of the method for each transient substructure
  of the NHIM.

  In this sense, the indicator function is the easier and more versatile method in
  general and it is also able to indicate the effects coming from the decaying parts of the
  NHIM. By relying on time delay functions exclusively one only loses some very fine
  details of the decay scenario.   However, with the indicator function, we only obtain
  information on the invariant sets that intersect the initial conditions chosen.

  We used an indicator function to display information
  on the localized chaotic invariant set of the system. This reminds us
  of the inverse chaotic scattering problem, where we use scattering functions
  and in particular also the time delay function as an indicator function to
  obtain information on the localized chaotic invariant set by asymptotic
  measurements. The only difference is that in scattering problems we use as the domain
  of the indicator function a subset of asymptotic initial conditions.
  In this sense, we move this domain to an infinite distance and for compensation, we have to subtract
  from the time delay the trivial asymptotic contributions which diverge to infinity.
  Otherwise, we have a perfect analogy. For general information on chaotic scattering see chapter
  6 in \cite{tel}, \cite{seoane} , and chapter 4 in \cite{reichl}; for examples of the use of the time delay
  function in the inverse chaotic scattering problem see \cite{jung2} and \cite{tapia}.
  Our important new addition to this general basic idea achieved in the present article is the special attention paid to tangential transient effects on unstable invariant subsets during their process of decay.

\section{Acknowledgments}

We thank DGAPA-UNAM for financial support under grant number IG101122 and
CONAHCyT for financial support under grant number 425854. FGM thanks the Faculty of Mathematics and Physics of the University of Ljubljana for their hospitality during the last stage of this work.

\end{document}